\documentclass[11pt, reqno]{article}
\usepackage{amsmath, amsthm} 

\usepackage{geometry}                
\usepackage{float}
\usepackage{graphicx}
\usepackage{subcaption}
\graphicspath{{Figures/}}
\usepackage{amssymb}
\usepackage{epstopdf}
\usepackage{braket}
\title{Boltzmann machines as thermal models for quantum systems}
\author{Steven Weinstein\thanks{sweinstein@pitp.ca}\\Perimeter Institute for Theoretical Physics\\University of Waterloo}
\date{May 24, 2021}                                           
\begin{document}
\maketitle
\abstract{We successfully model the behavior of two-spin systems using neural networks known as conditional Restricted Boltzmann Machines (cRBMs) which encode physical information in the properties of a thermal ensemble akin to an Ising model. The result gives local ``hidden'' variable models for product states and entangled states, including the singlet state used in the EPR-Bohm experiment. Bell's theorem is circumvented because the state of the system is dependent not only on the preparation but also on the measurement setup (the detector settings). Though at first glance counterintuitive, the apparent ``retrocausality'' in these models has a historical precedent in the absorber theory of Wheeler and Feynman \cite{WF45} and an intuitive analog in the simple AC circuit of an electric guitar.}  

\section{Introduction}

In 1935, Einstein, Podolsky and Rosen (EPR) gave a powerful argument that quantum mechanics is incomplete \cite{EPR35}.  Their central point is that if the predictions of quantum theory are correct, then the assumption of relativistic causality implies that there are facts about the world -- which EPR called `elements of reality' -- that are not reflected in the quantum state.\footnote{See \cite{Fine86} for a helpful reconstruction of the argument.} Thirty years later, Bell \cite{Bell64} proved that it is impossible to find a complete theory that satisfies certain seemingly natural constraints having to do with locality. The take-away for many has been that any theory describing quantum phenomena must be nonlocal in a way that would have been anathema to EPR, violating the spirit, if not the letter, of the constraints on causality enshrined in special relativistic theories.

In this paper, we offer a framework that provides a way of ``completing'' of quantum theory which shares much with classical statistical mechanics, and with a certain class of neural networks called Boltzmann machines \cite{HinSej83, AckHinSej85, Hin07}. Predictions in this framework are derived from the properties of a quasi-thermal ensemble, the boundary conditions of which are set by the classical configurations of the preparation and measurement apparatuses. Bell's argument is evaded because the assumption that the complete description of the system prior to measurement is \emph{independent} of the final measurement settings -- an assumption called Statistical Independence (SI) -- is violated.\footnote{SI is related to, but not the same as, measurement independence (MI).  The latter postulates that the detector settings are independent of the properties of the system being under measured.  It is also known as the ``free choice'' assumption, reflecting the idea that this is tantamount to the assumption that the settings can be freely chosen.} In fact, we will see in section \ref{section:nonlocal_boxes} that the framework generalizes in a natural way to allow for maximal violations of the Bell-inequality, giving a model that becomes both deterministic and nonlocal in a way that nevertheless does not allow superluminal signaling.\footnote{The model differs significantly from hidden-variable models like Bohmian mechanics \cite{Bohm52a, Bohm52b} and Nelson's stochastic mechanics \cite{Nel66, Nel85}, in that the stochasticity is not essential to avoiding the spectre of superluminal signaling.}

\section{Restricted Boltzmann Machines}

\begin{figure}[h] 
   \centering
   \includegraphics[width=0.65\textwidth]{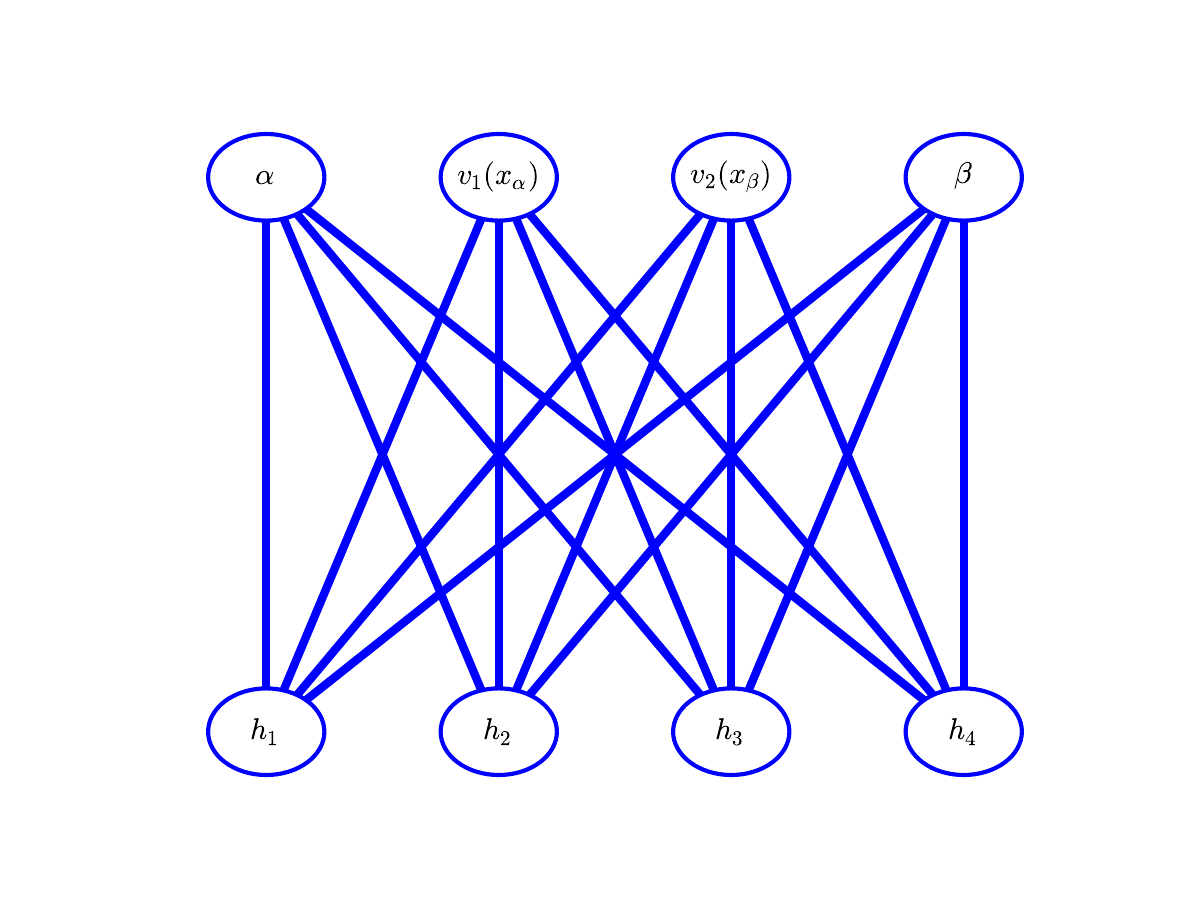} 
   \caption{Simple RBM for EPR with four hidden units}
   \label{fig:NNrbm_4}
\end{figure}

A Restricted Boltzmann Machine (RBM) \cite{Smo86} is a type of generative neural network with the topology of a bipartite graph (see Figure \ref{fig:NNrbm_4}).
For the models we work with, the nodes or units (as the `neurons' are called) are binary, like the spins in an Ising model. Edges are weighted connections between the units. Positive weights (excitatory connections) and negative weights (inhibitory connections) respectively encode correlations and anti-correlations between units. Each unit has a bias, which corresponds to an adjustable firing or activation threshold.  The \emph{visible} units (one layer of the graph) are intended to represent (binary) properties of the data, while the \emph{hidden} units (the other layer of the graph) encode correlations between these properties. 

For example, suppose we want our machine to learn a set of black-and-white images containing 16 x 16 = 256 pixels.  We will want to have 256 visible units $\mathbf{v} = (v_1, v_2,...,v_{256})$, one per pixel. The weights and biases will make some visible configurations more probable than others according to the \emph{ansatz} below. Training the machine involves systematically adjusting these weights and biases so that the machine assigns high probability to images like those in the data set. Thus a machine trained on 16 x 16 images of dogs and cats should generate images of dogs and cats, including \emph{but not limited to} those found in the training data. Because the hidden units $\mathbf{h} = (h_1, h_2,...,h_n)$ encode properties of dogs and cats, and the visible units manifest various combinations of these properties, sampling them will yield images of dogs and cats.

An RBM is a type of energy-based model.  To each configuration $(\mathbf{v}, \mathbf{h})$ there corresponds an energy
\begin{equation}
\label{eq:energy}
E(\mathbf{v},\mathbf{h}) = -\bigg( \frac{1}{2}\sum_{i=1}^m\sum_{j=1}^n w_{ij} v_i h_j + \sum_{i=1}^m c_i v_i  + \sum_{j=1}^n d_j h_j \bigg) .
\end{equation}
The probability of the configuration is given by the Boltzmann distribution for a system in thermal equilibrium at temperature $T$:
\begin{equation}
\label{eq:Boltz}
P({\mathbf{v, h}} ) = \frac{e^{-E(\mathbf{v, h})/T}}{Z}, \text{ where } Z=\sum_{\mathbf{v,h}} e^{-E(\mathbf{v,h})/T}.
\end{equation}
This will look familiar to any condensed matter physicist, as eqn. (\ref{eq:energy}) has the form of the Ising Hamiltonian.\footnote{This is no coincidence: the precursor to the Boltzmann machine was an associative memory model invented by physicist John Hopfield, who found his inspiration in the theory of spin glasses \cite{Hop82}.} \footnote{The idea of using an Ising model to represent physics which is temporally non-local is explored by Wharton in \cite{Whar14}.}
Just as in the Ising model, $w_{ij}$ is a weight, a connection strength between units $i$ and $j$, while $c_i$ and $d_j$ are 
the biases of visible unit $v_i$ and hidden unit $h_j$, respectively. (In the Ising model, these would correspond to the strength of 
the transverse magnetic field at each site.) 
The important difference between an RBM and an Ising model is the topology; whereas an Ising model typically has the connectivity of a lattice, an RBM is a bipartite graph, where each visible node is connected to every hidden node, and vice-versa. Also, whereas the Ising model units take values $\pm1$, the Boltzmann machine units take values $\{0,1\}$ by convention; one can move from one representation to the other by simply changing the biases. Unless otherwise indicated, we will henceforth work at fixed temperature and set $T=1$.

The probabilities of \emph{observable} quantities are the probabilities of various configurations of the visible units.  These are obtained by using the \emph{free energy} $\mathcal{F}$ associated with a visible configuration $\mathbf{v}$:
\begin{equation}
\label{eq:Free}
\mathcal{F}(\mathbf{v}) \equiv - \log \sum_{\mathbf{h}} e^{-E(\mathbf{v,h})}.
\end{equation}
This gives the effective energy needed to determine the marginal probability of all configurations that share the visible configuration $\mathbf{v}$.
The probability is then given by
\begin{equation}
\label{eq:Prob}
P(\mathbf{v}) = \frac{e^{-\mathcal{F}(\mathbf{v})}}{Z} \text{ with } Z=\sum_\mathbf{v'} e^{-\mathcal{F}(\mathbf{v'})}.
\end{equation}
$Z$ is the partition function, the functional form of which tells us how the free energy is apportioned amongst the various configurations, while its \emph{value} is a normalizing factor used to turn the exponentiated energies into probabilities.

RBMs are local in the same sense that the Ising model is local. The probability that a given unit is `on' depends only on the the state of the units to which it's connected. More specifically, 
\begin{equation}\label{updaterule}
P(v_i = 1) = \frac{1}{1 + e^{-{\Delta}E_{i}}}
\end{equation}
where
\begin{equation}
\begin{split}
{\Delta}E_{i} &= E_{v_i=0} - E_{v_i=1}\\
&= {c_i} +\sum_j w_{j}\,{h_j} .
\end{split}
\end{equation}
Thus the probability that $v_i$ is on depends only on the state of the hidden units $h_j$ to which it is directly connected. The bias $c_i $ and the weights $w_j$ are fixed parameters in the model.  

\section{RBMs and cRBMs for EPR}

An RBM is a useful way of learning and encoding a joint probability distribution.  
Consider the modern version of the EPR experiment \cite{EPR35} due to Bohm \cite{Boh51}.  It involves spin 
measurements on a pair of particles prepared in a maximally 
entangled `Bell' state, specifically two spin $\frac{1}{2}$ particles in the singlet state 
\begin{equation}
\label{eq:singlet}
\psi_{singlet} = \frac{1}{\sqrt{2}}\big(\ket{+-} - \ket{-+}) .
\end{equation}
One particle travels to station $A$ and the other to $B$.  At each station is a detector with two settings, each setting selecting a component of spin to be measured. The detector settings $\alpha \in \{a, a'\}$ and 
$\beta \in \{b, b'\}$ and measurement outcomes $x_\alpha \in \{+1, -1\}$ and $x_\beta \in \{+1, -1\}$ are 
two-valued, so we can treat them as binary random variables.  A single experimental trial $(\alpha,\beta,x_\alpha,x_\beta)$ can thus be represented as a binary vector $\mathbf{v}=(v_1, v_2, v_3, v_4)$, where $v_i \in \{0,1\}$.  This vector represents the values taken by the four visible units that make up one layer of the RBM.  (We map $+1$ to $0$ and $-1$ to $1$.)

We need to pick a sufficient number of hidden units $\mathbf{h}=(h_1, h_2, ... h_j)$ to encode the dependencies between the visible units. In \cite{SW17}, we used four hidden units, which was sufficient to allow the RBM to represent the quantum-mechanically predicted statistics of an EPR experiment to very high accuracy. The topology of the machine is that shown in Figure \ref{fig:NNrbm_4}.
Beginning with randomly chosen weights and biases, we 
trained the machine on a simulated data set from an EPR experiment using stochastic gradient descent on the maximum likelihood function to adjust the weights and biases, until the probabilities generated by the machine aligned with those in the data.  For 
example, if $\mathbf{v}=(1, 1, 1, 0)$, appears in 5.2\% of the trials, we would expect that the probability assigned to that vector by the 
machine would be appropriately close to 5.2\%, so that $P(\mathbf{v}) = .052$, where the probability is given by eqn. \ref{eq:Prob}.

In encoding the \emph{joint} probability $P(\alpha, \beta, x_\alpha, x_\beta)$, we have arguably encoded more information than necessary, if what we are trying to do is reproduce the predictions of quantum mechanics.  After all, quantum mechanics gives us only \emph{conditional} probabilities $P(x_\alpha,x_\beta | \alpha,\beta )$.  It tells us what we can expect to observe if we measure certain observables, but it does not tell us how often we can expect to measure those observables. Yet in specifying a joint probability distribution, we are effectively specifying both of these things.  At best, this means we are squandering the representational capacity of our model on quantities we're not interested in. So we now utilize a variant on an RBM called a conditional RBM (cRBM) to eliminate this redundancy.

\begin{figure}[h]
\centering
\begin{subfigure}{0.3\textwidth}
 \includegraphics[width=\textwidth]{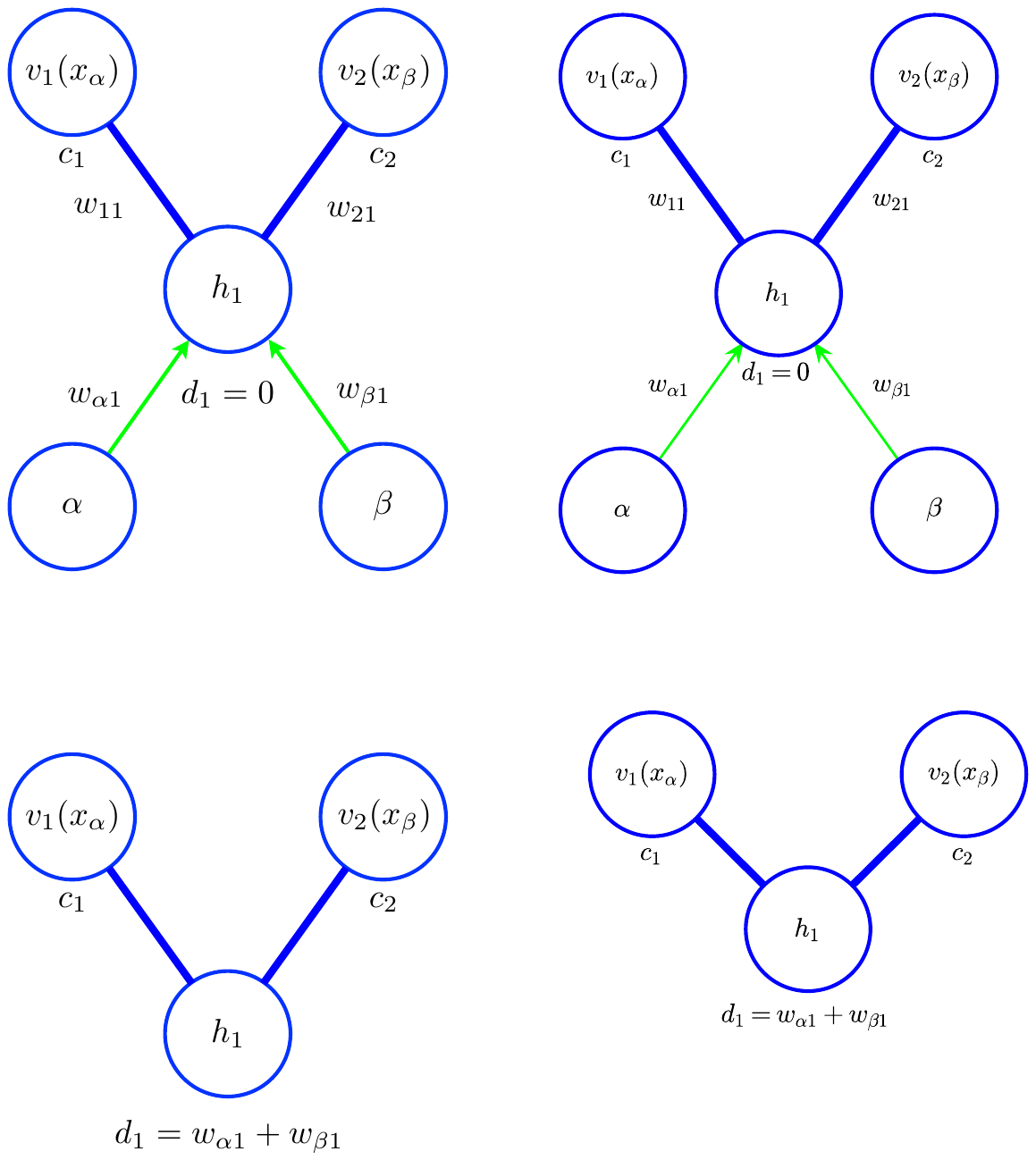}
 \caption{Conditional RBM}
 \label{fig:crbm_h1_a}
 \end{subfigure}
 \hspace{1in}
 \begin{subfigure}{0.3\textwidth}
 \includegraphics[width=\textwidth]{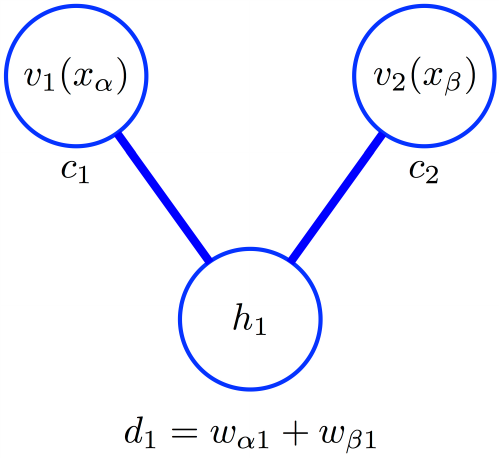}
 \caption{RBM with adjustable hidden unit bias}
  \label{fig:crbm_h1_b}
 \end{subfigure}
  \caption{Equivalent representations}
  \label{fig:crbm_h1}
\end{figure}

The essence of a cRBM is that it is able to encode conditional probability distributions directly, i.e., without deriving them from a joint probability distribution. This is done by allowing some or all of the biases and weights to vary according to the condition or conditions in question. In our cRBM, we allow each detector setting to influence the bias of the hidden units. (This is the part of the model that violates the statistical independence assumption used in Bell's theorem.) For the purposes of illustration, consider a cRBM with a \emph{single} hidden unit (figure \ref{fig:crbm_h1}).
We have two visible units $v_1$ and $v_2$ to represent the two outcomes $x_\alpha$ and $x_\beta$, and two additional conditional units $\alpha$ and $\beta$ to represent the detector settings. These are composite `one-hot' units, each of which consists of a set of binary units, one for each setting.  `One-hot' means that at any time, one and only one of the (binary) units is on and the others are off, so the unit as a whole is always on: it always has one ``hot'' (on) unit. As a result, the links between these conditional units and the hidden units are effectively one-way, directed links, since the hidden units have no effect on the conditional units. In fact, the conditional units simply act like adjustable biases on the hidden units. If the bias $d_1=0$, then one can view the machine shown in figure \ref{fig:crbm_h1_a} as an ordinary (non-conditional) RBM with an adjustable bias $d_1=w_{\alpha1} + w_{\beta1}$, as in figure \ref{fig:crbm_h1_b}. This is what we want: the detector settings are external inputs which have an effect on the parameters (here, the hidden unit biases) of the RBM model. Thus in the language of quantum foundations, the detector settings are correlated with the values of the hidden variables.

Training a cRBM is similar to training an RBM.  As with an RBM, the machine learns weights and biases so as to reproduce, in the activations of the visible units, the patterns found in the data. Let $\mathbf{u_1}$ correspond to the setting $\alpha$ of detector A, and $\mathbf{u_2}$ correspond to setting $\beta$ of detector B. For a given pair of settings $\mathbf{u} = (\mathbf{u_1}, \mathbf{u_2})$, the energy of the RBM is 
\begin{equation}
E(\mathbf{v}, \mathbf{u}, \mathbf{h}) = -\bigg( \frac{1}{2} \sum_{i=1}^m \sum_{j=1}^n w_{ij} v_i h_j + \sum_{i=1}^m c_i v_i  + \sum_{j=1}^n d_j  h_j  + \frac{1}{2} \sum_{j=1}^n\sum_{l=1}^2\sum_{k=1}^2 w_{jlk} u_{lk} h_j \bigg) .
\end{equation}
where $w_{jlk}$ refers to the weight connecting hidden unit $h_j$ to the $k$'th setting of one-hot unit $\mathbf{u_l}$. Then the conditional probabilities are simply
\begin{equation}
\begin{split}
P(\mathbf{v} | \mathbf{u}) &= \dfrac{\sum_h {e^{-E({\mathbf{v,u,h}})}}}{\sum_{v',u,h} {e^{-E({\mathbf{v',u,h}})}} }\\[2pt]
&= \dfrac{e^{-\mathcal{F}(\mathbf{v},\mathbf{u})}}{Z} .
\end{split}
\end{equation}
Training involves tailoring \emph{all} the weights, including the $w_{jlk}$ weights connecting the conditional units to the hidden units.  These weights encode the dependence of the stochastic properties of the hidden units on the detector settings.

What differentiates the cRBM from a regular RBM is that the units corresponding to the conditioning variables -- in this case, the detector settings -- are \emph{not} dynamical variables. Though the energy associated with a configuration is in part a function of the weights connecting the conditioning units to the hidden units, the sum over the configurations in the partition function involves only configurations in which the detectors and the preparation (the input state) have some specific value. Thus there are no probabilities assigned to detector settings or states. The only probabilities generated by the cRBM are conditional probabilities, conditional on the states and detector settings, just as in ordinary quantum mechanics.

\section{Results}

We constructed a variety of cRBM models, and learned the weights and biases using simple stochastic gradient descent, as described in, e.g., \cite{Hin12}.
All of the models have \emph{three} conditioning units, one for each of two detectors (as above) and an additional one corresponding to the choice of input state. They vary in the number of possible settings for each detector (2 or 8), the number of states (singlet only, or singlet plus two product states), and the number of hidden units.  The intent was not only to model one particular experiment, the EPR experiment, but also to explore the ability of cRBMs to model generic two-particle systems with arbitrary preparations and detector settings.   

\begin{figure}[h] 
   \centering
   \includegraphics[width=0.8\textwidth]{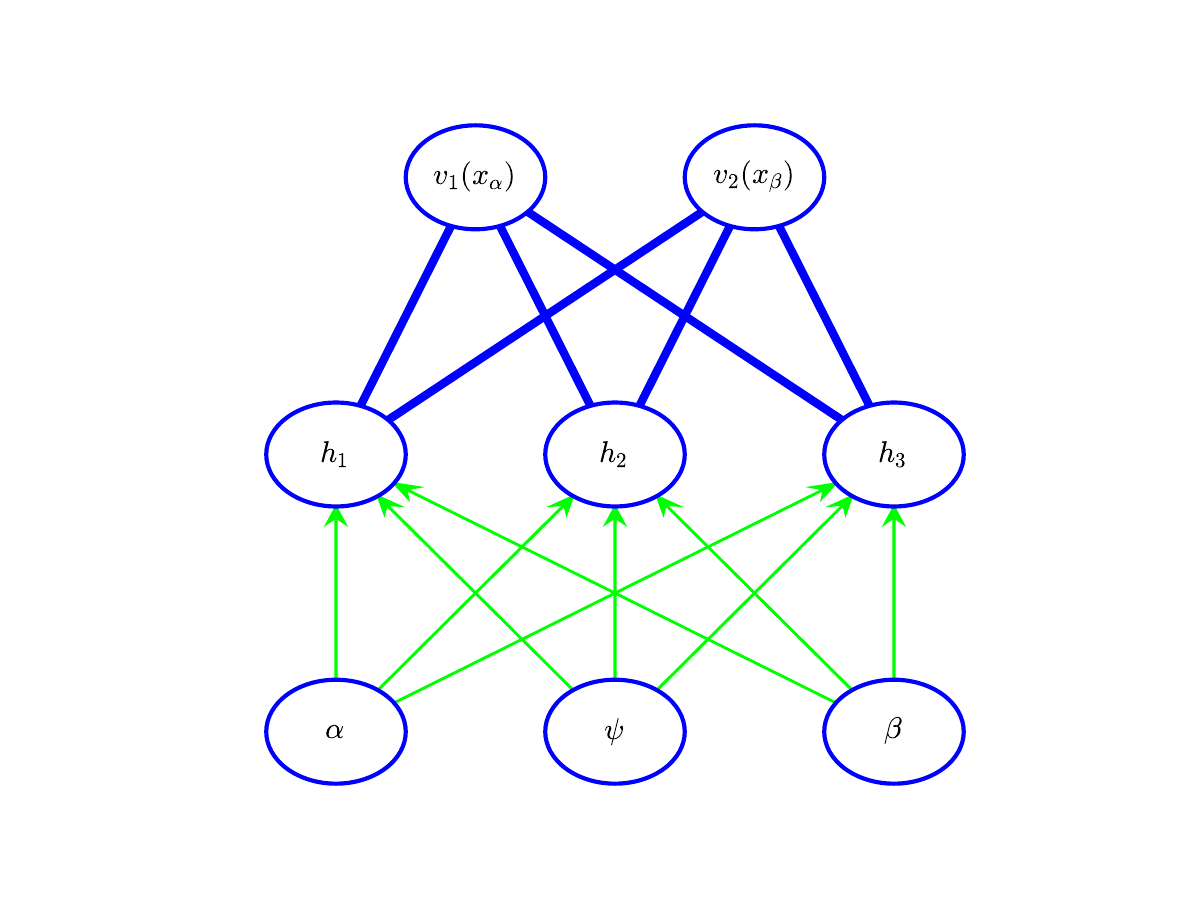} 
   \caption{Conditional RBM for EPR with three hidden units}
   \label{fig:NN_crbm122_h3}
\end{figure}

\begin{figure}[h] 
   \centering
   \includegraphics[width=\textwidth]{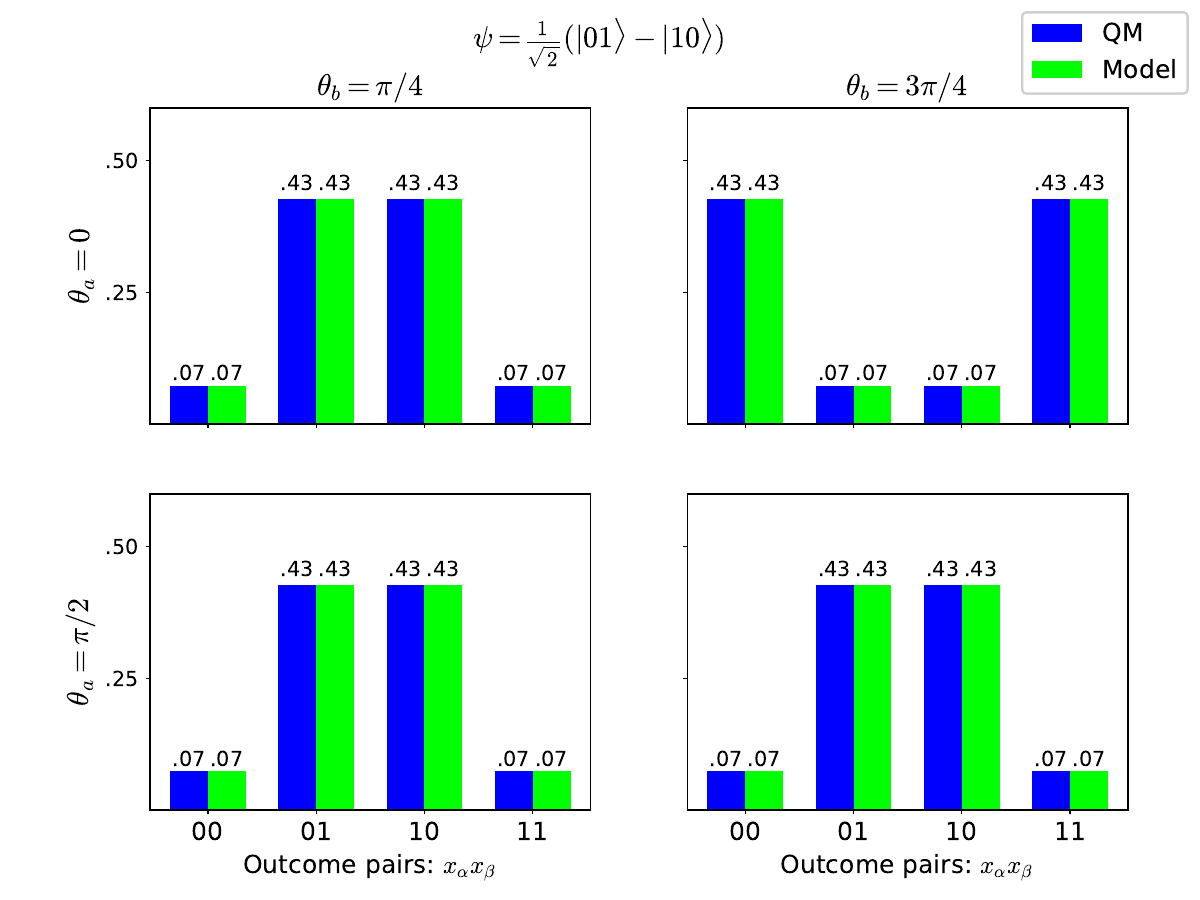} 
   \caption{Predictions for cRBM model with three hidden units}
   \label{fig:Bar_122st2_h3}
\end{figure}

A simple cRBM model with three hidden units easily reproduces the success of the (non-conditional) RBM model with four hidden units discussed previously. The topology of the model is shown in figure \ref{fig:NN_crbm122_h3}, and the results after training are shown (to two decimal places) in figure \ref{fig:Bar_122st2_h3}. Here $\psi$ is the singlet state (\ref{eq:singlet}) and the settings $\alpha$ and $\beta$ can each take one of two values corresponding to different components of spin.

\begin{figure}[h] 
   \centering
   \includegraphics[width=\textwidth]{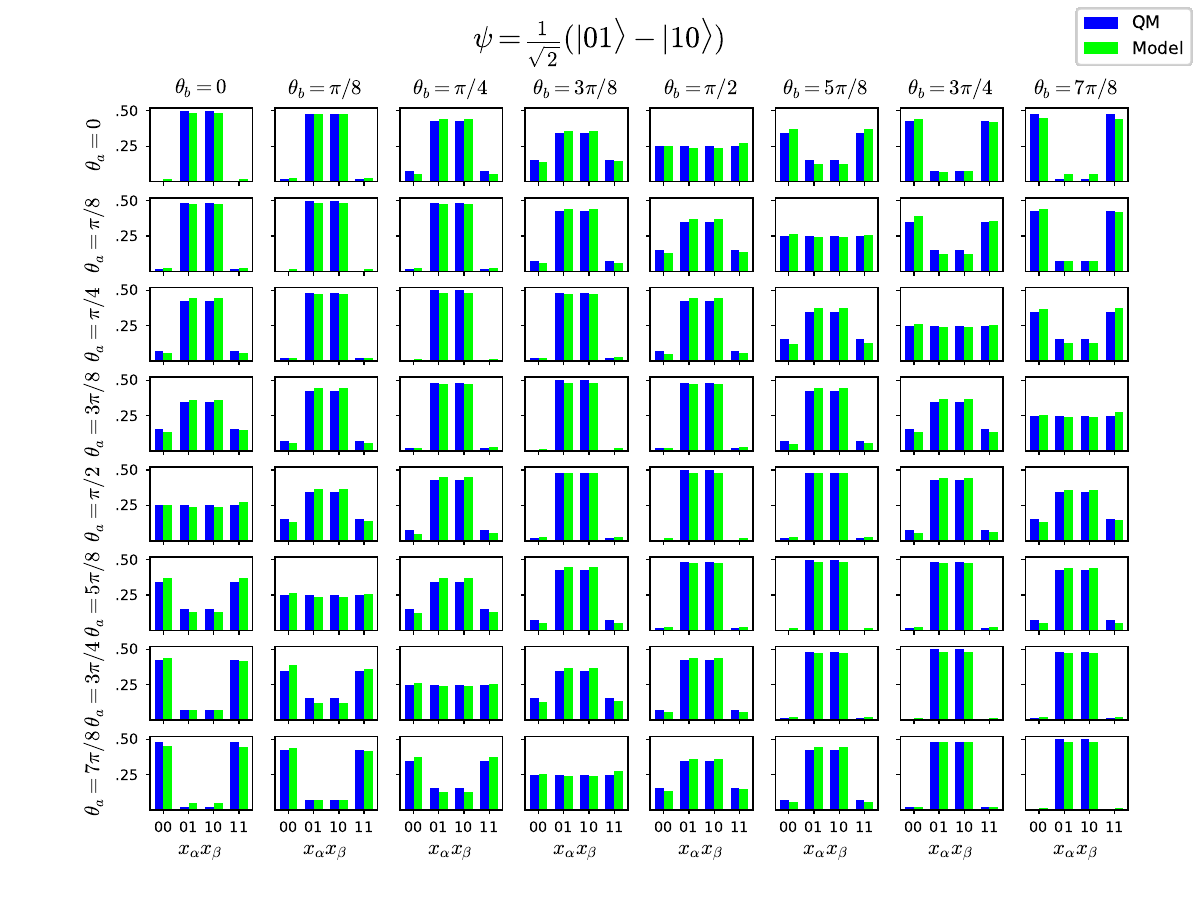} 
   \caption{Predictions for cRBM model with eight hidden units: Singlet state}
   \label{fig:Bar_188st2_h3}
\end{figure}

\begin{figure}[h] %
   \centering
   \includegraphics[width=\textwidth]{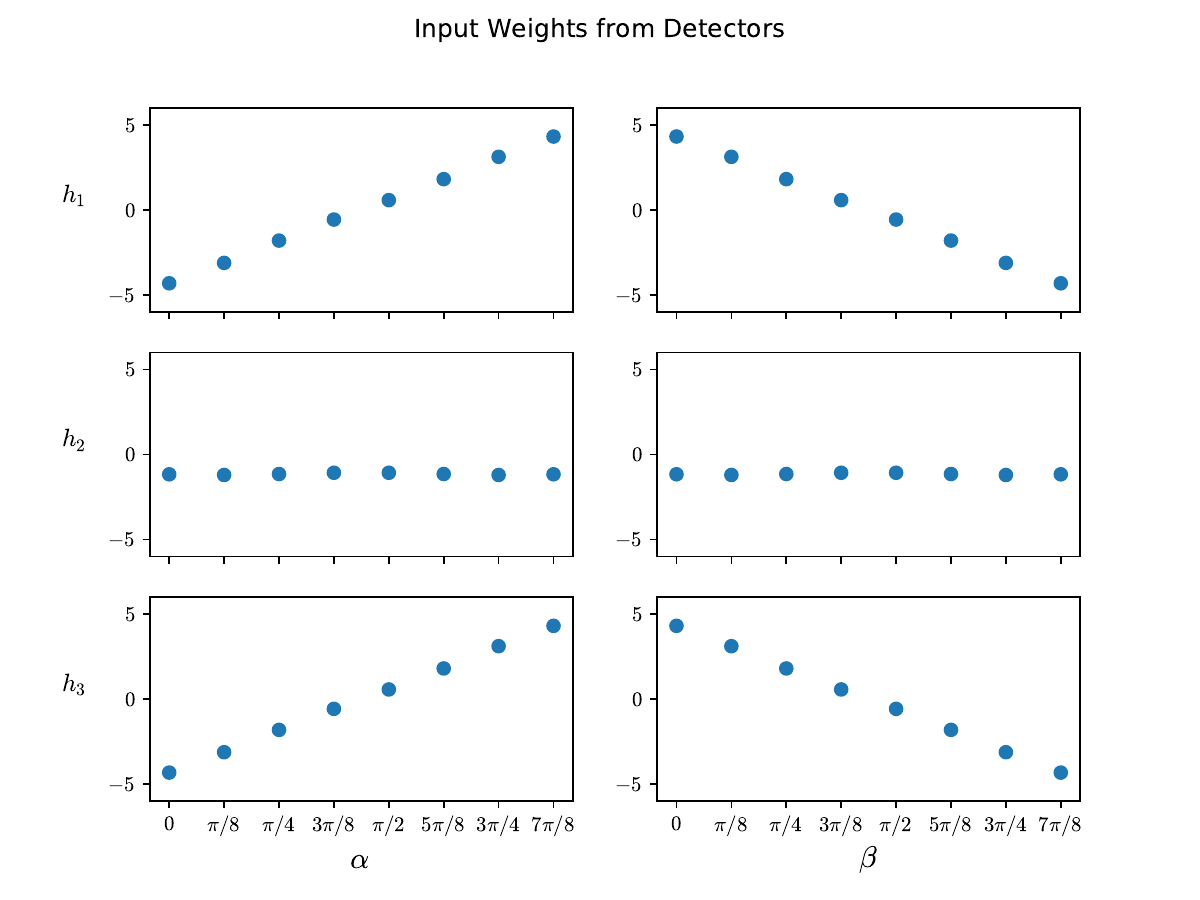} 
   \caption{Input weights for detectors A and B}
   \label{fig:Plot_settings_crbm188_h3}
\end{figure}

Aiming to generalize the result, we constructed a model with \emph{eight} possible settings between $0$ and $7\pi/8$ for each detector, still for a single entangled state.  The topology is the same as before. The results are shown in figure \ref{fig:Bar_188st2_h3}.
Even with eight settings, three hidden units were sufficient to give very good results. Taking a look at the directed links from the units modeling the detectors and the three hidden units in figure \ref{fig:Plot_settings_crbm188_h3}, we see a more or less linear relationship between detector setting and weight for the connections between both of the detectors and two of the three hidden units, while the other unit displays only subtle variation for each of the detectors.

The next step was to try to model the same measurements on systems prepared in different quantum states.  In addition to the singlet (eqn \ref{eq:singlet}), we chose the two product states $\ket{+-}$ and $\ket{-+}$.   Given that a general quantum state of two particles requires seven real numbers to specify (four complex numbers $c_i$ such that $\sum_{i=1}^4 |c_i|^2 = 1$) , it is not that surprising that we needed eight hidden units to get the level of accuracy seen in figures \ref{fig:Bar_388st0_h8}, \ref{fig:Bar_388st1_h8} and \ref{fig:Bar_388st2_h8}.
 \begin{figure}[!tbp] 
   \centering
   
  \begin{subfigure}[c]{\linewidth}
   \includegraphics[width=0.9\textwidth]{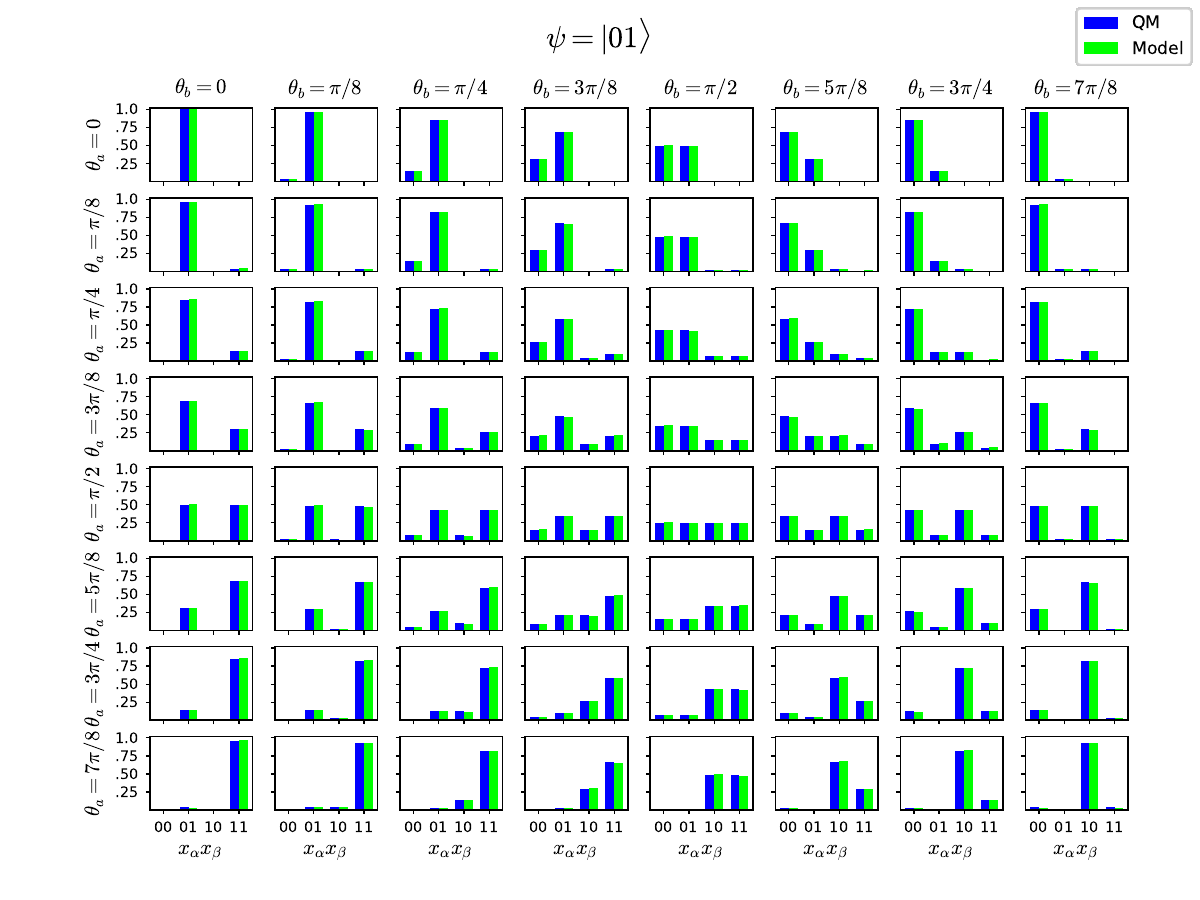} 
   \caption{Predictions for cRBM model with eight hidden units: State $\ket{+-}$}
   \label{fig:Bar_388st0_h8}
\end{subfigure}

\begin{subfigure}[c]{\linewidth}
   \includegraphics[width=0.9\textwidth]{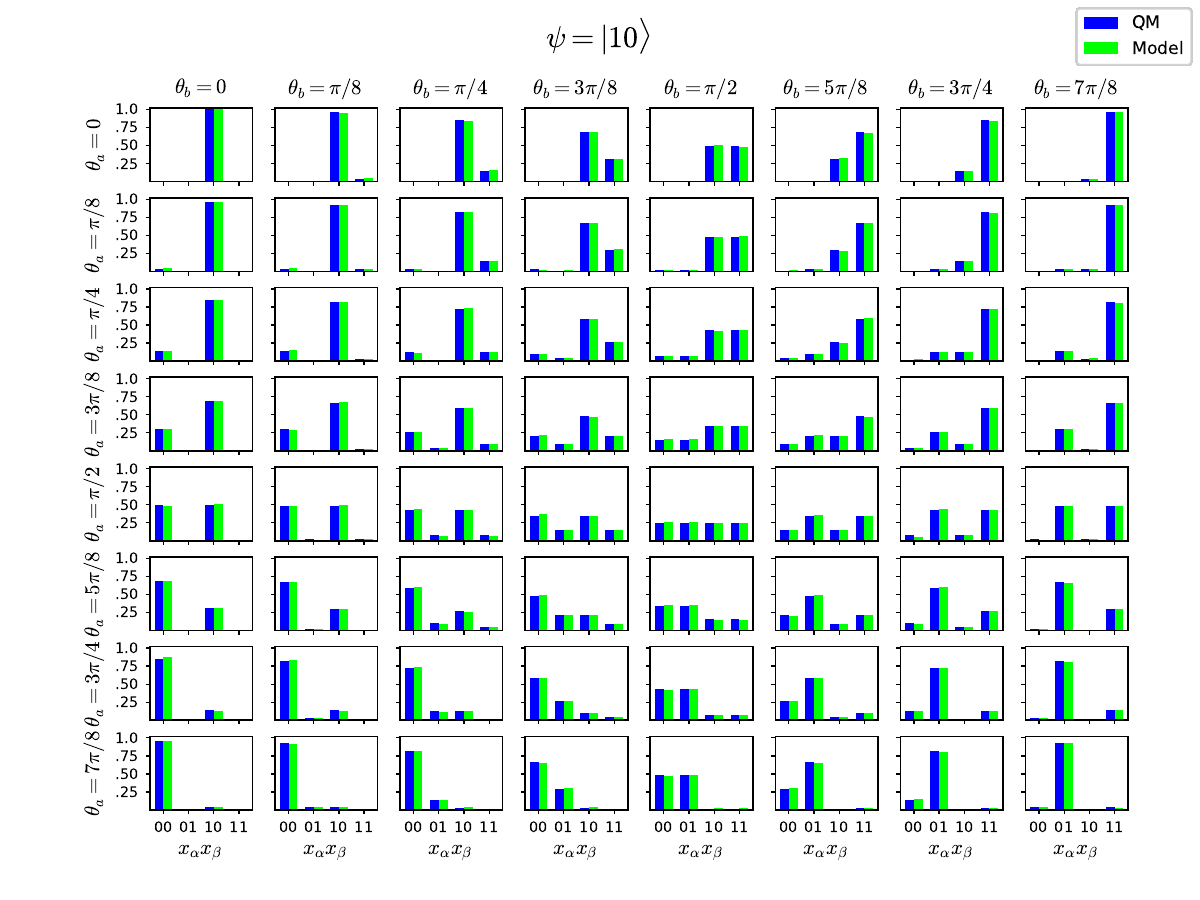} 
   \caption{Predictions for cRBM model with eight hidden units: State $\ket{-+}$}
   \label{fig:Bar_388st1_h8}
\end{subfigure}
\end{figure}

 \begin{figure}[!tbp] 
 \ContinuedFloat
   \centering
\begin{subfigure}[c]{\linewidth}
   \centering
   \includegraphics[width=0.9\textwidth]{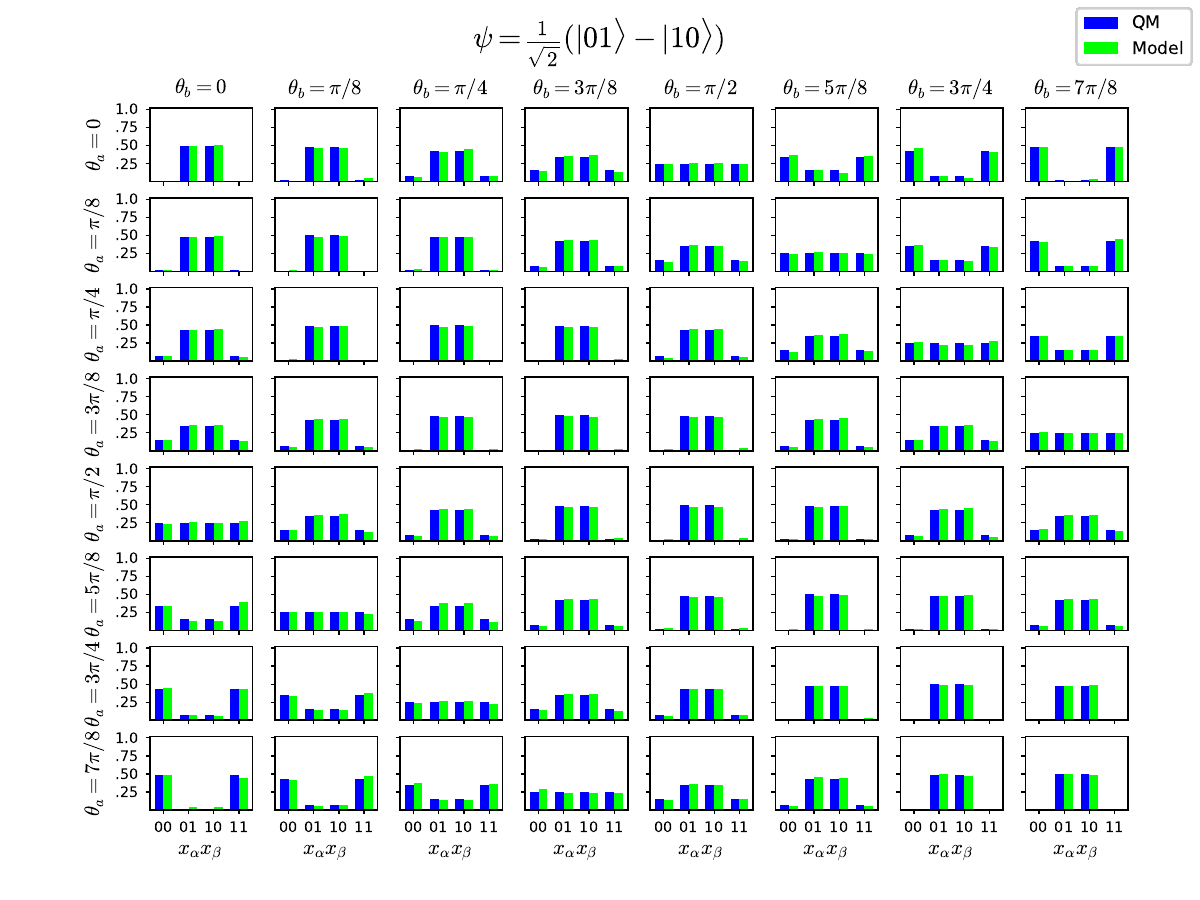} 
   \caption{Predictions for cRBM model with eight hidden units: Singlet state}
   \label{fig:Bar_388st2_h8}
 \end{subfigure}
\caption{Predictions for cRBM model with eight hidden units: Three states}
\end{figure}

\section{Bell's theorem and cRBM models}

The cRBM models provide a template for a ``completion'' of quantum mechanics.  Bell's theorem is widely understood to show that no \emph{local} theory can reproduce all the predictions of quantum mechanics.  So how does our model get around this? 

Bell works in a framework where each choice of state (each choice of values for the dynamical variables or `beables', hidden or not) leads to probabilities for the two possible outcomes of spin measurements on an entangled pair of spin-$\frac{1}{2}$ particles. If the model that gives rise to the probabilities satisfies certain properties, it must satisfy the Bell-CHSH inequality \cite{CHSH69}: that is Bell's theorem.  For appropriate choice of detector settings -- appropriate choice of components of spin to be measured -- quantum mechanics predicts, and experiment appears to corroborate \cite{CHSH69, FC72, ADR82}, that this inequality is violated. This shows that a theory that satisfies the conjunction of Bell's `locality' condition and in which the detector settings and dynamical variables are independent cannot reproduce the predictions of quantum mechanics. The cRBM models reproduce the quantum predictions but do not violate Bell's theorem because although they are `local', they do not satisfy SI, the condition of statistical independence. We have previously dubbed this \emph{nonlocality without nonlocality} \cite{SW09}.

Bell's analysis is intended to characterize a wide variety of candidate replacements or completions of quantum mechanics. The supposition is that in any given trial the system is in some state $\lambda$ drawn from a space of states $\Lambda$ defined by the model, and that to each such state there corresponds a prediction (in general, stochastic) about the results of measurements on each of the two particles.
Naturally, the probability that a given state $\lambda$ will be realized is going to depend in part on the conditions $c$ that generate the pair of particles. Thus we speak of the the conditional probability $P(\lambda | c)$.  This much conditional dependence of the $\lambda$ on external boundary conditions is uncontroversial, and is assumed by Bell.  

Now, each state $\lambda$ implies probabilities $P(x_\alpha x_\beta | \alpha, \beta, \lambda) $ which yield an expectation value $-1 \leq E(\alpha, \beta | \lambda) \leq +1$ for the product of the outcomes $x_\alpha, x_\beta \in \{+1, -1\}$:
\begin{equation}
\label{eq:expect_a}
E(\alpha, \beta | \lambda) \equiv \sum_\lambda \sum_{\alpha, \beta} x_\alpha x_\beta  P(x_\alpha x_\beta | \alpha, \beta, \lambda) \end{equation}
The expectation value is then the weighted sum:
\begin{equation}
\label{eq:expect_b}
E_c(\alpha, \beta) = \sum_\lambda E(\alpha, \beta | \lambda) P(\lambda | c)
\end{equation}
Expressions of the form $S = \sum_{\alpha,\beta} | \pm E(\alpha, \beta)| $ must take a value $S \leq 4$, since each expectation value has a maximum value of $+1$ and a minimum value of $-1$.   In particular,
\begin{equation}
\label{CHSH}
S = |(E(a,b) + E(a',b) + E(a,b') - E(a',b'))| \leq 4 .
\end{equation}

For spatially separated measurements, one might suppose that, given a particular state, the probabilities factorize as follows:
\begin{equation}
\label{eq:locality}
P_\lambda(x_\alpha, x_\beta | \alpha, \beta)  = P_\lambda(x_\alpha, \alpha) * P_\lambda(x_\beta | \beta) .
\end{equation}
This is Bell's `locality' condition, later distinguished by him from the `local causality' condition, which is the conjunction of locality and the additional condition of statistical independence (SI),
\begin{equation}
\label{eq:SI}
P(\lambda | c) = P(\lambda | c, \alpha, \beta),
\end{equation}
which implies that the detector settings do not constrain the state $\lambda$. Together, locality and SI imply the CHSH-Bell inequality: $S \leq 2$.  But if the conditions $c$ correspond to the preparation of a singlet state, and the detector settings are such that $(a, a', b, b') = (0, \pi/2, \pi/4, 3\pi/4)$, then quantum mechanics predicts that $S = 2\sqrt{2}$.  

\section{Physical models with future boundary conditions}

\begin{figure}[!tbp] 
   \centering
   \includegraphics[width=0.8\textwidth]{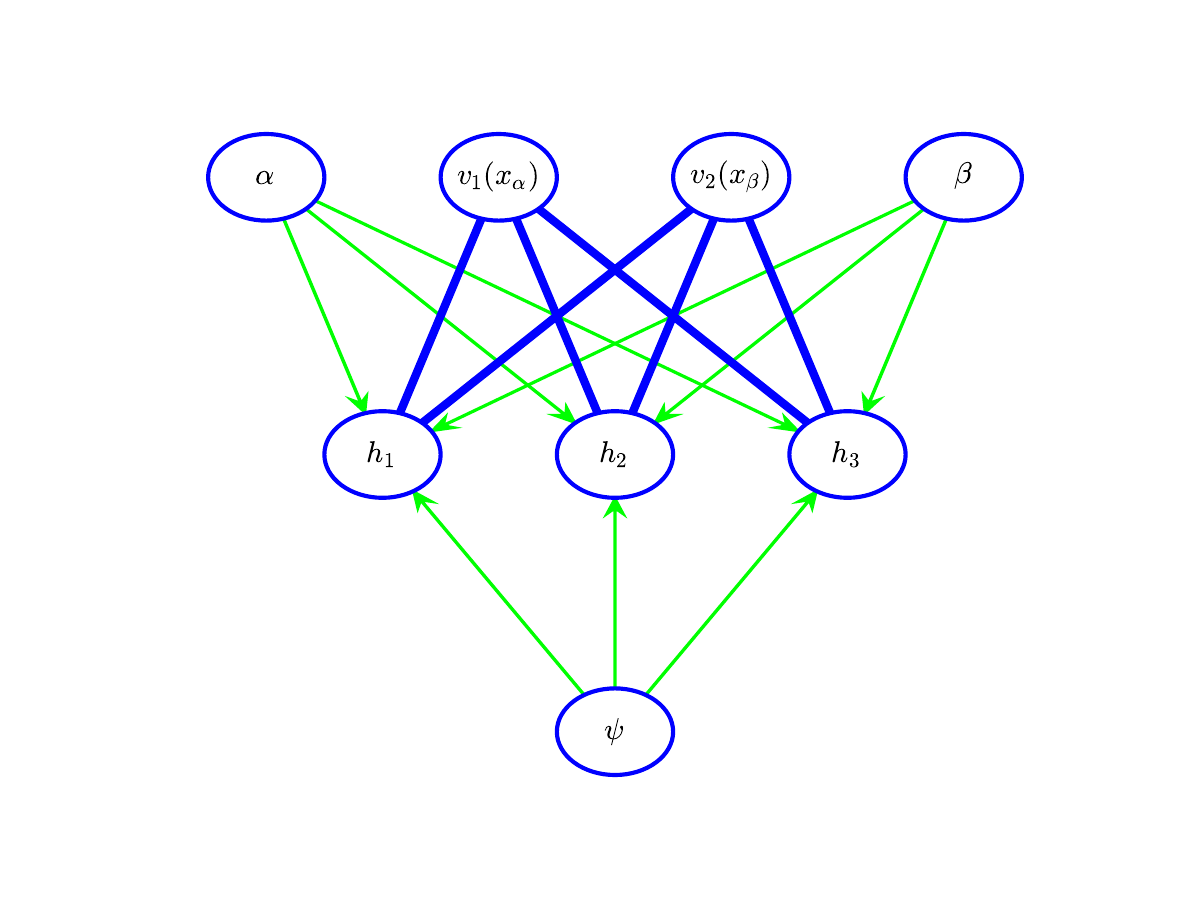} 
   \caption{Conditional RBM for EPR with three hidden units}
   \label{fig:NNa_crbm122_h3}
\end{figure}

In the cRBM model, the state of the hidden units is a function of both the input conditions (which we refer to as $\psi$), and the detector settings $\alpha$ and $\beta$. The machine is in a state of quasi-thermodynamic equilibrium at temperature $T=1$ with future and past boundary conditions, so that it forms a kind of spacetime ensemble.  Arrayed in spacetime, with time running from bottom to top, we can view the machine depicted in figure \ref{fig:NN_crbm122_h3} in the manner depicted in figure \ref{fig:NNa_crbm122_h3}.

Models in which ``freely specifiable'' future boundary conditions constrain the state of the system earlier in time are often called `retrocausal' \cite{Cos78, Cram86, ACE15, PW17, LP17}, and have been the subject of much recent discussion \cite{WharArg20}. Yet the idea is arguably not that novel. The most prominent example of such a theory might seem to be ordinary classical mechanics formulated in the Lagrangian manner. Though the usual use of the Lagrangian formulation is to derive the equations of motion from arbitrary initial and final system configurations, one can also specify particular initial and final configurations and derive a trajectory -- not necessarily unique -- from the principle of least action.
What one cannot do, of course, is to specify initial and final configurations \emph{and} velocities.  In that case, the problem is overdetermined, meaning that in general there will be no solution at all. But specifying partial data at the initial and final times leads to a well-defined, if not strictly well-posed (because of non-uniqueness) problem. And given that we typically regard configurations of physical systems as ``freely specifiable'', Lagrangian models would superficially appear to be candidates for the retrocausal appellation. 

A model which more closely meets the retrocausal criterion, one in which past and future boundary conditions \emph{uniquely} determine the present, is the reformulation of Maxwell's electrodynamics known as the Wheeler-Feynman absorber theory \cite{WF45, Davies71, Davies72}, which was indeed inspired by Lagrangian mechanics. This theory treated the electromagnetic field as a useful fiction, and derived the standard behavior of charged particles under the assumption that all radiation -- if it actually existed -- would be absorbed in the future. Thus in this theory, there is no source-free radiation, and the initial and final charge distributions determine the behavior of charges in the intermediate region.  The primary drawback is that the universe does not appear to be configured such that all radiation is ultimately absorbed, since the expansion renders it largely transparent  \cite{Hogarth62}.

\begin{figure}[!tbp]
\centering

\begin{subfigure}[c]{.4\linewidth}
 \includegraphics[width=\textwidth]{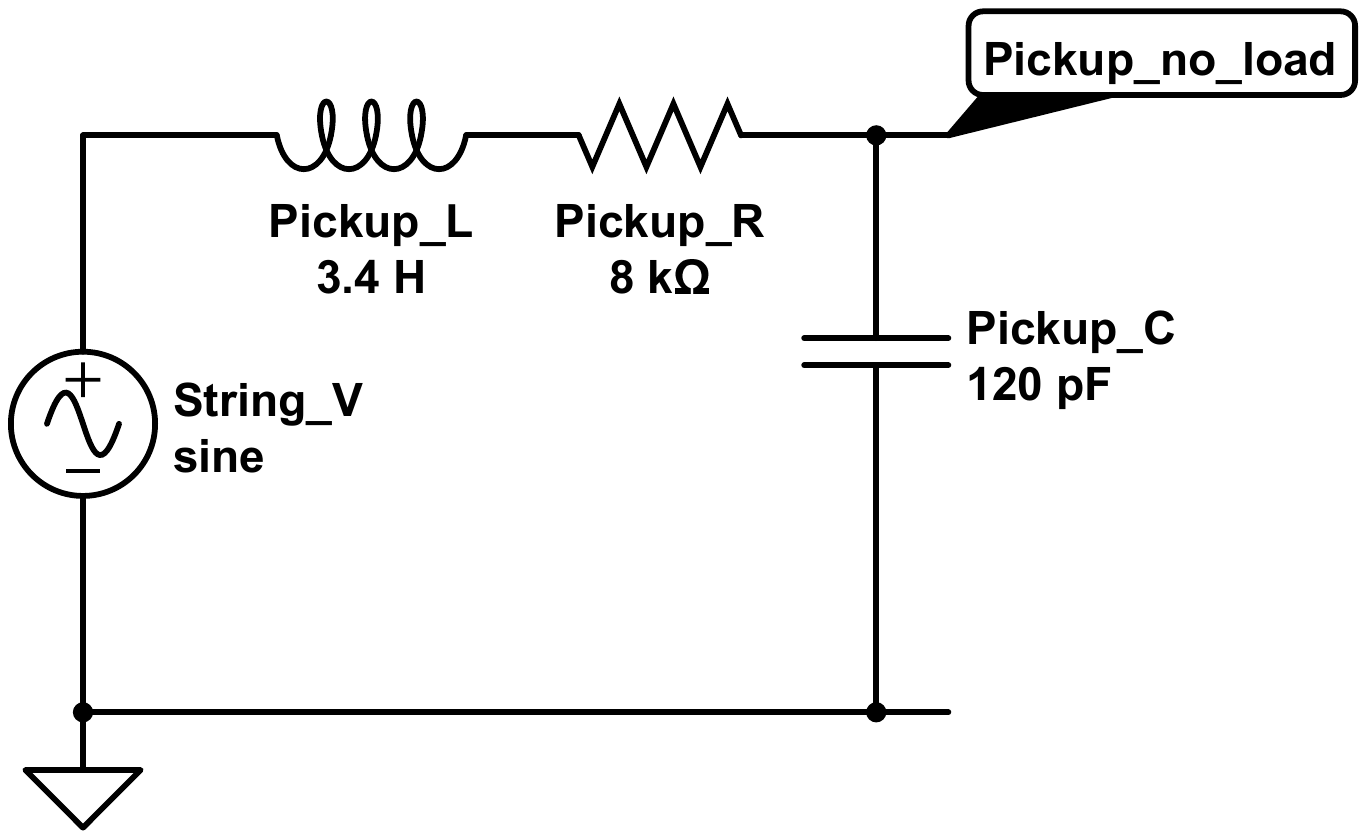}
 \caption{Electric guitar pickup driven by vibrating string. No load.}
 \label{fig:noload_schem}
 \end{subfigure}
 \hfill
 \begin{subfigure}[c]{.45\linewidth}
 \includegraphics[width=\textwidth]{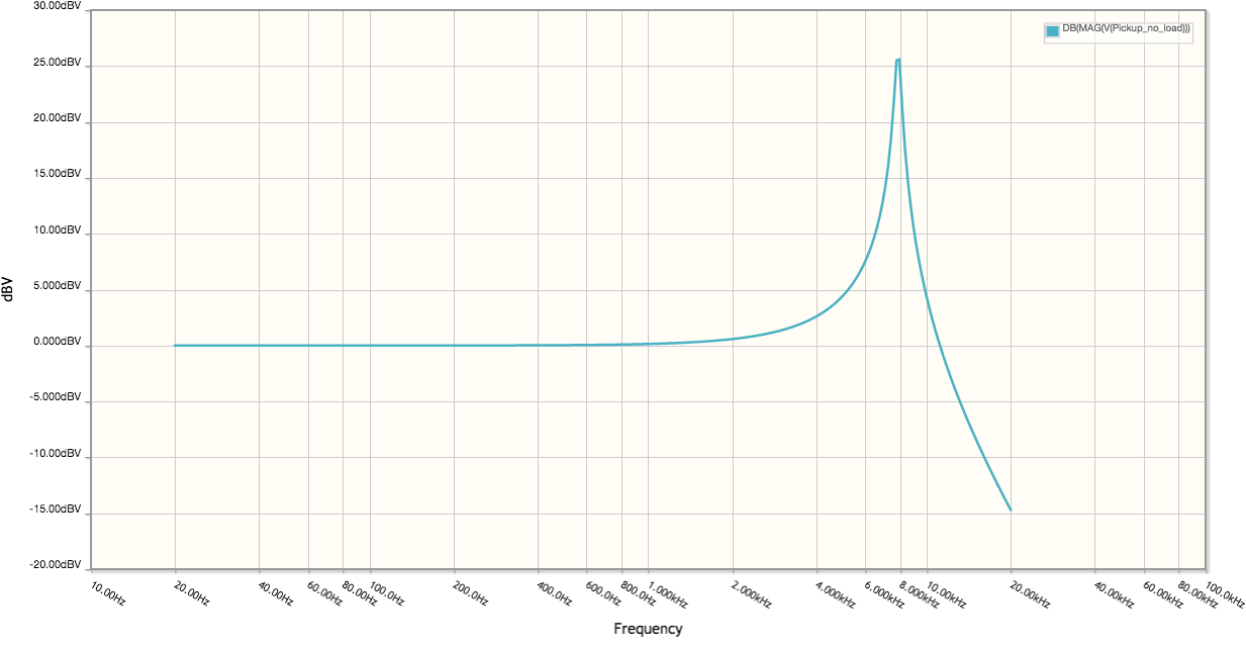}
 \caption{Frequency response of the unloaded pickup}
  \label{fig:noload_freq}
 \end{subfigure}
 
  \begin{subfigure}[c]{.5\linewidth}
 \includegraphics[width=\textwidth]{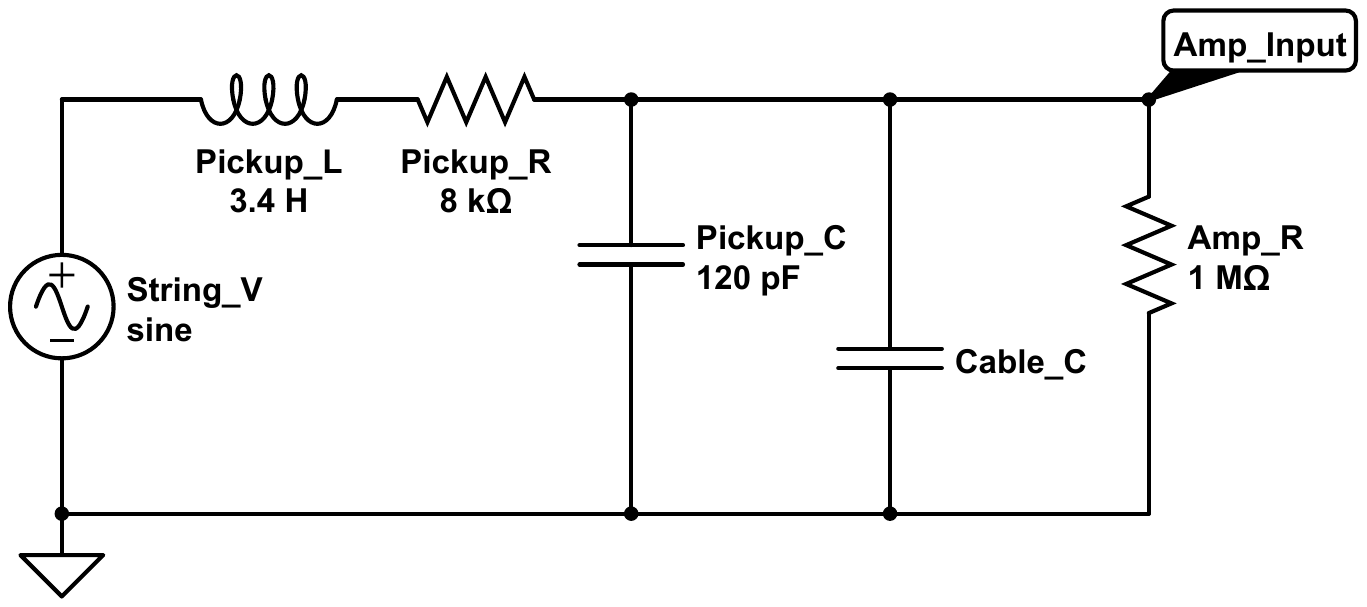}
 \caption{Pickup connected to amplifier with cables of varying capacitance. Sampled at amplifier.}
 \label{fig:cable_schem2}
 \end{subfigure}
\hfill
\begin{subfigure}[c]{.45\linewidth}
 \includegraphics[width=\textwidth]{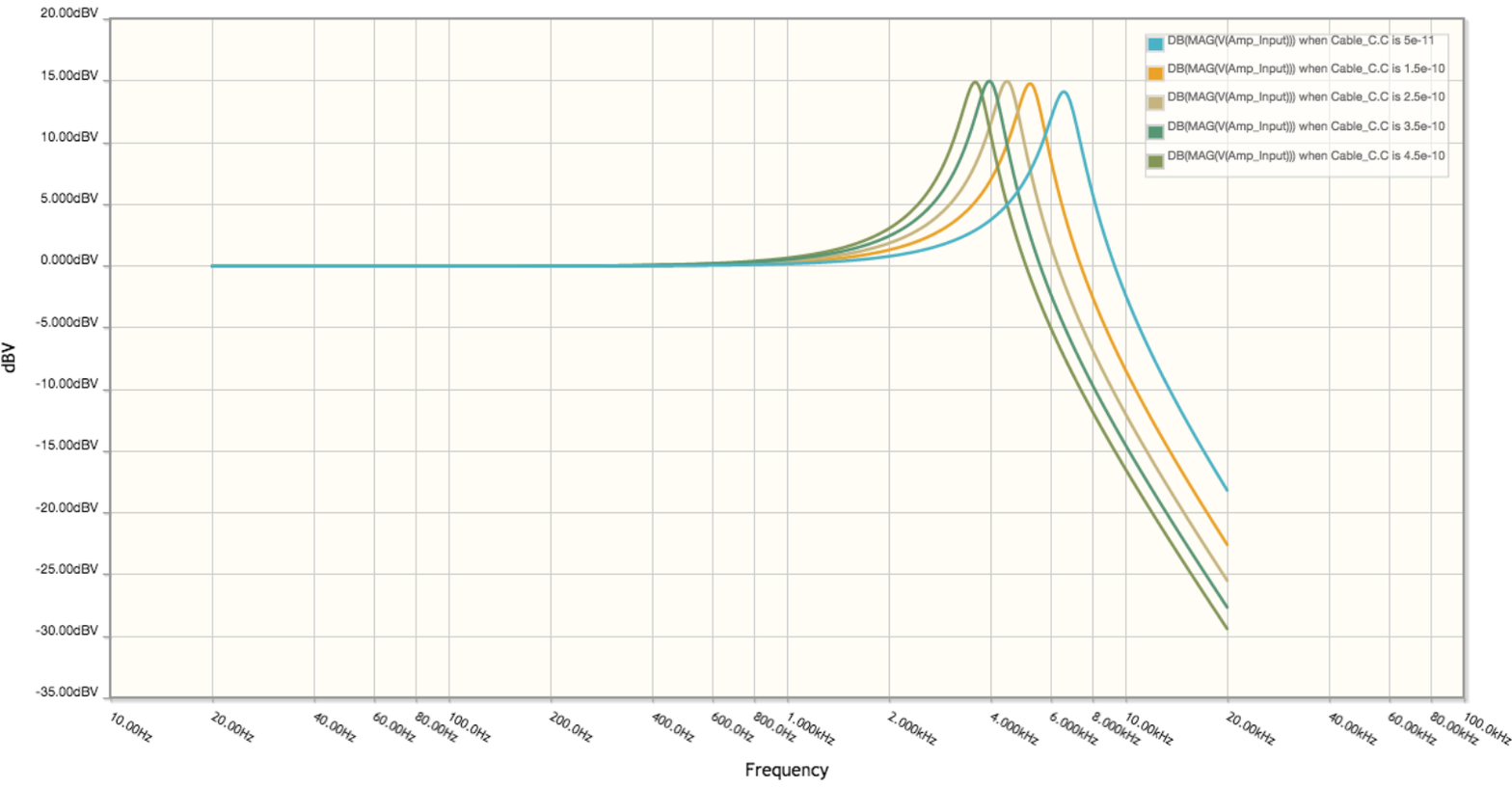}
 \caption{Frequency response at amplifier}
  \label{fig:cable_freq2}
 \end{subfigure}
 
\begin{subfigure}[c]{.5\linewidth}
 \includegraphics[width=\textwidth]{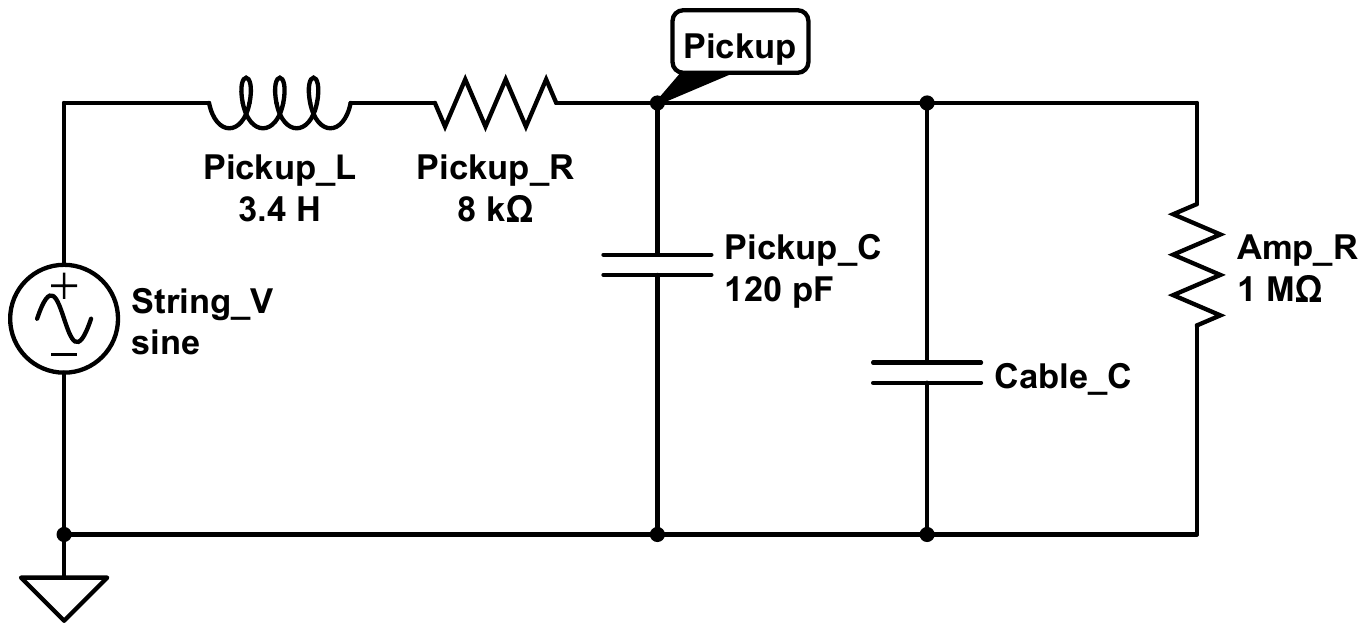}
 \caption{Pickup connected to amplifier with cables of varying capacitance. Sampled at pickup.}
 \label{fig:cable_schem1}
 \end{subfigure}
\hfill
\begin{subfigure}[c]{.45\linewidth}
 \includegraphics[width=\textwidth]{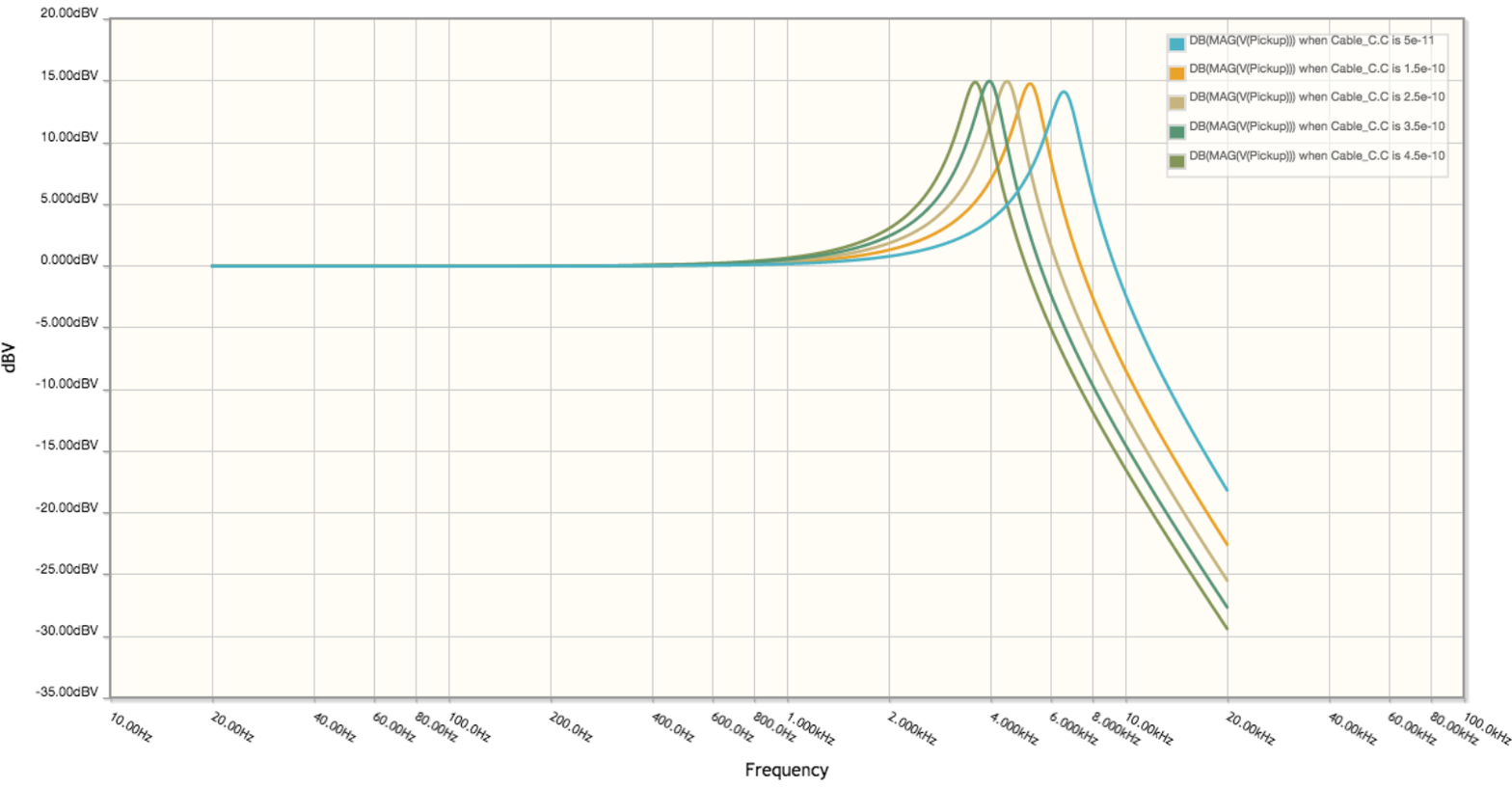}
 \caption{Frequency response at pickup}
  \label{fig:cable_freq1}
 \end{subfigure}
 
  \caption{Frequency response of electric guitar pickup under varying capacitive loads}
  \label{fig:pickup}
\end{figure}

An everyday example of a dynamical system governed by past and future boundary conditions is the most important instrument of the 20th century, the electric guitar. At its heart is a simple, damped resonator called a `pickup', which consists of a magnet within or underneath a coil of wire. The magnet magnetizes the steel strings, and the strings' oscillation drives the circuit by inducing an alternating current in the coil via electromagnetic induction, in accord with Faraday's law. Thus the pickup serves as a transducer, converting the mechanical oscillation of the string into an electrical oscillation. But the pickup's frequency response is not flat, so it colors the signal: it is a second-order low-pass filter. Just like a mechanical oscillator, the pickup has a resonant frequency at which it is most easily driven (see figure \ref{fig:pickup}).

The frequency response of the pickup depends not only on its internal properties (most importantly the inductance of the coil), but on the properties of the objects \emph{receiving} the signal, later in time. The resonant frequency of the pickup $f_r = \sqrt{1/LC}$ is a function not only of the inductance $L_p$ and capacitance $C_p$ of the coil, but also the capacitance of the cable connecting the guitar to the amplifier.  

Under no external load, the pickup (figure \ref{fig:noload_schem}) reacts to the input signal as depicted in figure \ref{fig:noload_freq}.  
When the guitar is plugged into an amplifier, the frequency response changes, depending on the load. In figure \ref{fig:cable_freq2}, we see the frequency response at an amplifier with input impedance of 1 M$\Omega$ using cables with capacitance ranging from 50 pf to 450 pf. And what is especially significant for our purposes is that the behavior at the pickup is the same as the behavior at the amp. The cable affects the way the input signal is altered, but it does so in an \emph{effectively} retrocausal way, as one can see by examining the frequency response at the pickup (figure \ref{fig:cable_freq1}). The signal proceeds unchanged from the pickup through the cable to the amplifier, even though the properties of the cable affect the signal. The signal seems to ``know'' about the cable before it gets there.

Thus in this simple AC circuit, the signal is affected not only by the source (the preparation) but by the receiver (the measuring device). Viewed this way, it is very much analogous to the cRBM models. Of course, there is no \emph{true} backward causation here, because changes made later in the chain take a finite, if miniscule, time to propagate back to the pickup. But this timescale is irrelevant for understanding the behavior of the system, since the signal of interest consists of oscillations whose period (at minimum $5 * 10^{-5}$ seconds for the highest audible frequency) is several orders of magnitude larger than what one might call the relaxation time of the system as a whole (around $10^{-8}$ seconds).  In other words, the time it takes for a change at the amplifier input to propagate back to the pickup is 5000 times less than the time of one cycle of the highest audible frequency in the signal. Thus it is effectively instantaneous, so that changes made downstream at the cable or amplifier influence the signal upstream in the pickup itself.   

\section{Statistical independence, measurement independence, and ``fine-tuning''}

Models of EPR correlations that allow interdependence of some sort between the states $\lambda$ and the detector settings $\alpha, \beta$ have been studied under various rubrics.  Models that violate SI are sometimes called \emph{retrocausal} as they postulate a dependence of the states at earlier times on the detector settings at the time of measurement. Models that violate what is sometimes called \emph{measurement independence} postulate that the conditional dependence goes the other way around, so that the seemingly-freely chosen detector settings are at least partially constrained by the prior states \cite{Brans88, Hall10, Hall16, BG11, Hooft03, Hooft14, Handsteiner17}. These are often called \emph{superdeterministic}, and are said to violate the ``free will'' assumption.  Of course it's possible to have the dependence running in both directions, in case the model has a well-defined joint probability over detector settings and states. 

Wood and Spekkens \cite{WS15} showed that all models that postulate some sort of mutual dependence of states and settings must be ``fine-tuned'' if they are to rule out superluminal signaling. The motivation for this is the idea that if the hidden state is influenced by the choice of setting at A, and if it in turn affects the outcome at B, then it will take a very special choice of model parameters (e.g., the weights and biases in the cRBM model) in order to preclude the possibility of faster-than-light signaling from A to B.  They prove that any such model which does \emph{not} permit signaling can be turned into a model which does allow signaling by altering the model parameters.  Essentially, this means that the causal/topological structure of the model is by itself insufficient to preclude signaling.

The argument about fine-tuning is intended to suggest the unnatural or even conspiratorial nature of SI-violating models. Almada \emph{et al.} \cite{Almada15} have countered that the various symmetry constraints that we commonly appeal to as constraints on our theories would appear to represent a kind of fine-tuning on the Wood and Spekkens characterization. Though further study is needed, it is indeed likely that it is the symmetries in the weights and biases of the cRBM model that preclude superluminal signaling \cite{Adlam18a}. These are quite evident already in the simple model shown in figure \ref{fig:Plot_settings_crbm188_h3}.

\section{Nonlocal boxes as the $T \rightarrow 0$ limit} \label{section:nonlocal_boxes}

The cRBM models exhibit interesting behavior in the $T \rightarrow 0$ limit.  As noted above, probabilities are generated via a Boltzmann distribution at $T=1$.  Thus we have a quasi-thermal ensemble at fixed temperature.  Given that this is a four-dimensional (spacetime) ensemble bounded by two times, rather than a spatial volume, it seems plausible that the physical counterpart of this temperature in our application is another constant, Planck's constant $\hbar$, this one having dimensions of action rather than energy. Indeed, the temperature in our cRBM model is playing a role similar to that played by $\hbar$ in the path integral formalism. One might regard the study of the $T \rightarrow 0$ limit as the analogous study of the $\hbar \rightarrow 0$ limit in quantum mechanics or any other theory in which $\hbar$ plays a fundamental role.

\begin{figure}[h] 
   \centering
   \includegraphics[width=0.65\textwidth]{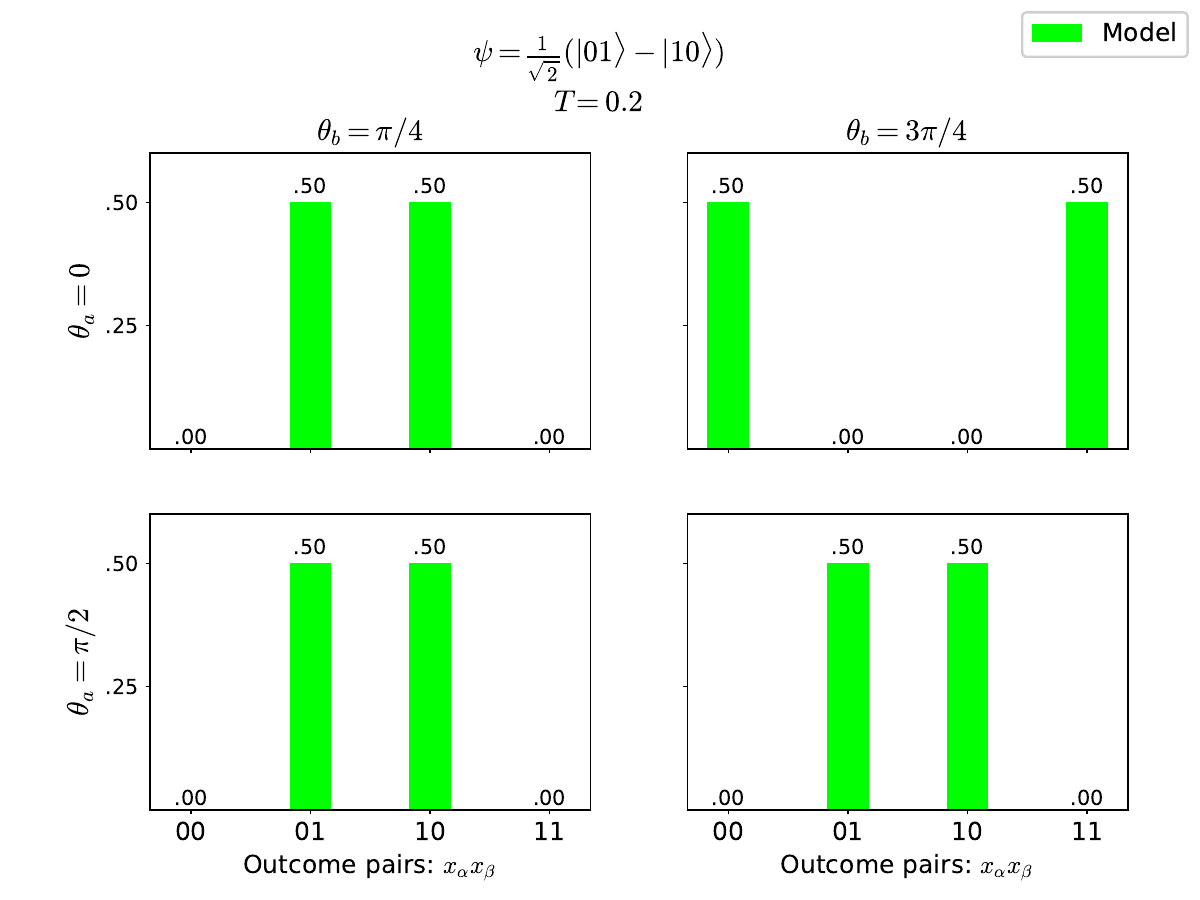} 
   \caption{cRBM model with three hidden units at $T=0.2$ approximates a PR box}
   \label{fig:Tbar_122st2_h3_Tpt2}
\end{figure}

Whether or not one regards the temperature as analogous to Planck's constant, the $T \rightarrow 0$ limit is interesting because it reproduces ``PR boxes'' \cite{PR94}, hypothetical deterministic devices the behavior of which yields maximal violation of the Bell-inequality ($S = 4$) while not allowing superluminal signaling. They are maximally nonlocal boxes. In figure \ref{fig:Tbar_122st2_h3_Tpt2} we see the results of the model for $T=0.2$, well above $T=0$ but nevertheless sufficient to give PR boxes with high accuracy.
Thus the cRBM model has a well-defined deterministic limit which is maximally nonlocal.

\section{Conclusion}

By training conditional Restricted Boltzmann machines, we have constructed a family of stochastic hidden-variable models for particle pairs, both entangled and not, for a variety of measurement settings. The methods used are quite general, and the patterns exhibited in the weights and biases suggest that one might be able to generalize the model to arbitrary detector settings and state preparations, and then to arbitrary numbers of particles. The biggest barrier to this would seem to be the need for exponential growth in the number of hidden units in order to accommodate the rich phenomenology attendant to the exponential growth of the Hilbert space dimension in the traditional quantum mechanical representation. Issues of practicality are indeed at the forefront of the attempt to use neural networks to learn particularly salient properties of large-scale quantum systems \cite{Carleo17}.

For our purposes, however, the proof of principle points in another, more interesting direction. That is, it points to the fact that a hidden-variable theory that depends on future boundary conditions associated with the settings of measuring devices \emph{may not tell us anything at all} about the physical world for systems that do not have any sort of effective measurement in their future. In cosmological settings, this is the norm, since the future lightcone of an arbitrary point in an expanding universe is transparent; there is no interaction or absorption. This in turn suggests the possibility of a new sort of limit on the applicability of quantum mechanics: quantum mechanics applies \emph{only} to systems which eventually undergo a measurement-like interaction. \emph{A fortiori}, it would not apply to systems with a transparent future lightcone.

The possibility of such a limit to the applicability of quantum mechanics might be a clue as to why the zero-point energy predicted by traditional quantum field theory does not seem to gravitate.  In a model in which detector settings co-determine the hidden variables in their past via the setting of the bias of those hidden units, the lack of any such interaction would mean that the bias is entirely indeterminate, and that the model, perfectly adequate for situations in which a detector-like interaction occurs in the future, is silent on the physics of particles or fields which never interact. Given that the vast majority of points in space at the present time appear to have an empty future lightcone (no future absorber), this gives us plenty of leeway to rethink the physics of those points, which constitute the bulk of the contribution to the zero-point energy, and thus the cosmological constant associated with traditional quantum field theory.

\section*{Acknowledgements}
Thanks to many for helpful and interesting discussions, among them Miles Blencowe, Lucien Hardy, Roger Melko, Rahul Sarpeshkar, Lee Smolin, Rob Spekkens and Ken Wharton.


\end{document}